\documentclass{webofc}
\usepackage[varg]{txfonts}   
\pdfmapfile{=rtxr.map}%
\pdfmapfile{=rtxmi.map}%
\pdfmapfile{=txsy.map}%
\begin{document}
\title{%
Search for
astrophysical PeV gamma rays
from point sources
with \textit{Carpet-2}
}
%
%

\author{%
\lastname{D.\,D.\,Dzhappuev}\inst{1} \and
\lastname{I.\,M.\,Dzaparova}\inst{1,2} \and
\lastname{E.\,A.\,Gorbacheva}\inst{1} \and
\lastname{I.\,S.\,Karpikov}\inst{1} \and
\lastname{M.\,M.\,Khadzhiev}\inst{1} \and
\lastname{N.\,F.\,Klimenko}\inst{1} \and
\lastname{A.\,U.\,Kudzhaev}\inst{1} \and
\lastname{A.\,N.\,Kurenya}\inst{1} \and
\lastname{A.\,S.\,Lidvansky}\inst{1} \and
\lastname{O.\,I.\,Mikhailova}\inst{1} \and
\lastname{V.\,B.\,Petkov}\inst{1,2} \and
\lastname{K.\,V.\,Ptitsyna}\inst{1} \and
\lastname{V.\,S.\,Romanenko}\inst{1} \and
\lastname{G.\,I.\,Rubtsov}\inst{1} \and
\lastname{S.\,V.\,Troitsky}\inst{1}\fnsep\thanks{Presenter at VLVNT-2018;
             \email{st@ms2.inr.ac.ru}} \and
\lastname{A.\,F.\,Yanin}\inst{1} \and
\lastname{Ya.\,V.\,Zhezher}\inst{1}
}

\institute{%
Institute for Nuclear Research of the Russian Academy of
Sciences,\\
60th October Anniversary prospect 7A, 117312 Moscow, Russia
\and
Institute of Astronomy, Russian Academy of Sciences, Moscow, 119017
Russia
}

\abstract{%
Early results of the search for $E_{\gamma}>1$~PeV cosmic photons from
point sources with the data of {\it Carpet--2}, an air-shower
array equipped with a 175~m$^{2}$ muon detector, are presented.
They include 95\%~CL upper limits on PeV photon fluxes from stacked
directions of high-energy IceCube neutrino events and from four predefined
sources, Crab, Cyg~X-3, Mrk~421 and Mrk~501. An insignificant excess of
events from Mrk~421 will be further monitored.
Prospects of the
use of the upgraded installation, {\it Carpet--3} (410~m$^{2}$ muon
detector), scheduled to start data taking in 2019, for searches of
$E_{\gamma}>100$~TeV photons, are
briefly discussed. }
\maketitle
\section{Introduction}
\label{sec:intro}
The data on astrophysical photons with energies above $\sim 0.1$~PeV are
presently very scarce, though their observation or non-observation is very
important for understanding many astrophysical and particle-physics
phenomena. The cosmic gamma rays of $E_{\gamma }\sim(0.1-10)$~PeV produce
electron-positron pairs on the Cosmic Microwave
Background~\cite{Nikishov}. This process limits their mean free path to
the size of the Galaxy. If no new physics is assumed, every single
detected astrophysical photon of this energy is certainly Galactic. This
makes PeV photons very useful in constraining unknown astrophysical
sources, notably those of high-energy neutrinos detected by IceCube (see
e.g.\ Refs.~\cite{Murase-gamma, OK-ST-gamma} for more details). In
addition, any observation of an air shower induced by a primary gamma ray
of that high energy would strongly constrain hypothetical models with
tiny, though presently not excluded, deviations from the Lorentz invariance
\cite{LIV1}. Contrary, any observation of PeV photons coming from an
extragalactic source would suggest new physical phenomena, for instance,
the axion-photon mixing (see e.g.\ Ref.~\cite{ST-axion-rev} for a review).
Sadly, no astrophysical PeV photons have been firmly detected yet.

This work presents early results of the search for cosmic photons with
energies $E_{\gamma}\gtrsim 1$~PeV with the \textit{Carpet--2} extensive
air shower array at the Baksan Neutrino Observatory of INR RAS.
\section{The method}
\label{sec:method}
\textit{Carpet--2} detects the electromagnetic component of extensive air
showers by means of its surface array of scintillator detector stations,
while the muonic component is detected by the underground scintillator
detector. The installation is described in detail
elsewhere~\cite{0902.0252, Carpet-2, 1511.09397}. For the present work, we
use the data recorded in 1999--2011 (total of 3080 live days, 115821
events which passed the quality cuts described in Ref.~\cite{CarpetIC}).
The number of $E \ge 1$~GeV muons in the 175~m$^{2}$ underground detector,
$n_{\mu}$, together with the shower size $N_{e}$, which is reconstructed
from surface-detector data, arrival direction and time represent the basic
information we use for each event. Since the showers initiated by primary
gamma rays are muon-poor compared to hadronic ones, $n_{\mu}$ helps to
distinguish the two kinds of primaries. Since both $n_{\mu}$ and $N_{e}$
scale roughly linear with the primary energy, we use their ratio,
$n_{\mu}/ N_{e}$, as the observable to separate gamma-like events, while
$N_{e}$ is used to estimate the primary energy.

Nowadays, PeV gamma-ray astronomy is in high demand after the IceCube
discovery of PeV neutrino events~\cite{IC}, because neutrinos and photons
are expected to be co-produced in hadronic interactions. However, at the
time the installation started data taking, its primary goal was to study
the structure of hadronic showers and not to search for elusive
photon-induced events. Therefore, unfortunately, the condition $n_{\mu}>1$
was included in the trigger, so that muon-poor ($n_{\mu}=0$ and 1) events
were not recorded. This criterion reduced considerably the efficiency of
the detector for gamma-ray primaries at lower energies; however, at
$E_{\gamma} \gtrsim $PeV, even photon-induced showers normally have
$n_{\mu} \ge 2$.

To estimate quantitatively the efficiency of the detection of primary
photons and the precision of the reconstruction of EAS parameters,
Monte-Carlo (MC) simulations were performed with the CORSIKA
package~\cite{CORSIKA} (v.~7.4003), using the hadronic-interaction models
QGSJET-01c~\cite{QGSJET-01} and FLUKA-2011.2c~\cite{FLUKA1}. Simulated
showers were thrown randomly on the installation, whose response was
simulated with additional dedicated MC procedures. For the resulting
artificial events, observable parameters (geometry, $N_{e}$ and $n_{\mu}$)
were reconstructed in the same way as it is done for the data. The
reconstruction efficiency for primary photons integrated over the field of
view (zenith angles $\theta \le 40^{\circ}$)
is $\sim 16\%$ at $E_{\gamma}\sim 1$~PeV and quickly reaches $>95\%$ at
$E_{\gamma}\sim 5$~PeV. The arrival directions of $E_{\gamma}>1$~PeV
photons are reconstructed with the accuracy of 1.8$^{\circ}$ (68\% CL),
2.65$^{\circ}$ (90\% CL).

\begin{figure}
\centering
\sidecaption
\includegraphics[width=0.62\textwidth,clip]{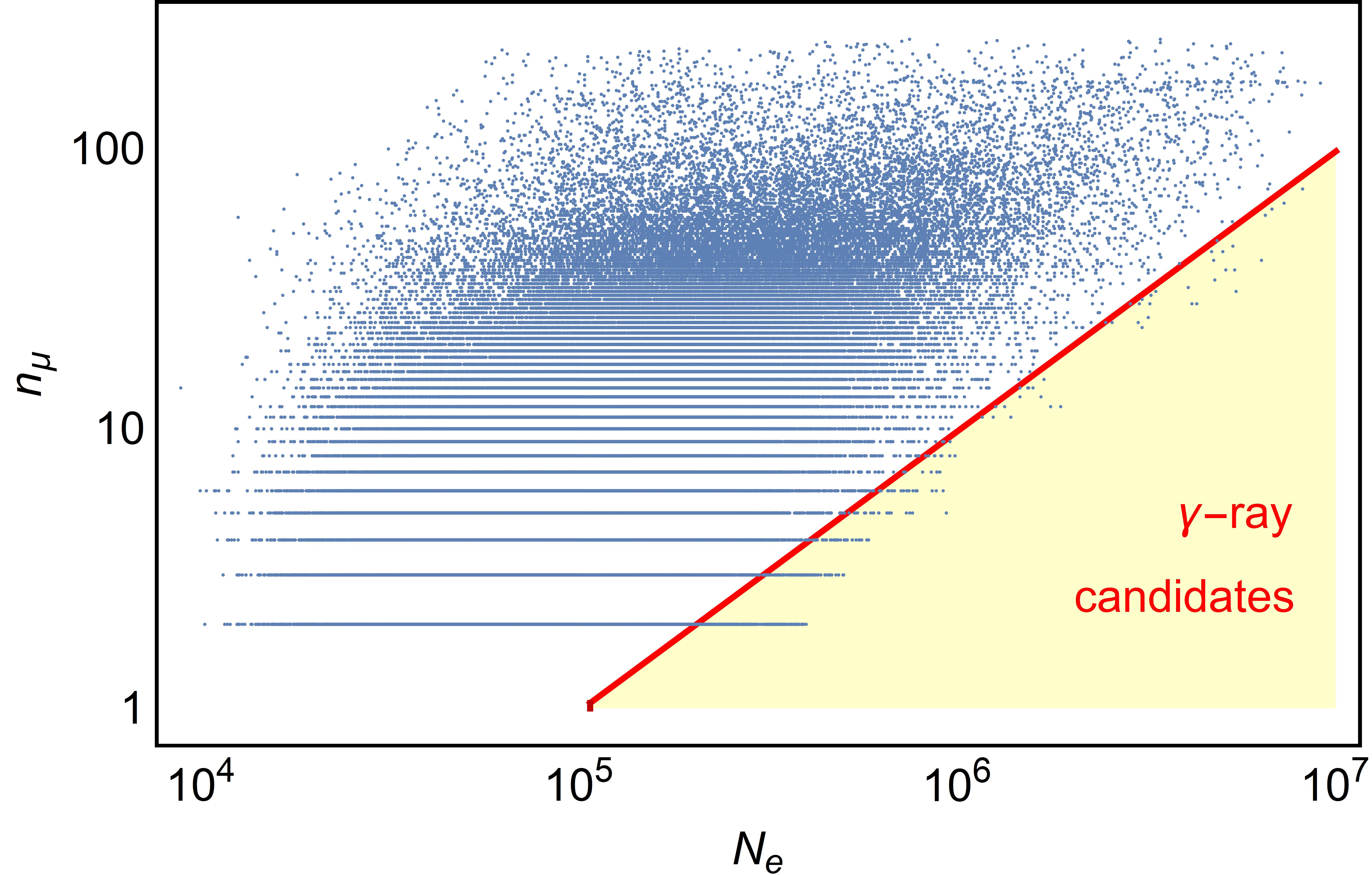}
\caption{Selection of photon candidate events on the $N_{e} - n_{\mu}$
plane. Blue dots represent all Carpet-2 events which passed the quality
cuts. Photon candidates (muon-poor events) are in the shaded region. See
the text for more details.}
\label{fig:cand-cuts}
\end{figure}

Based on the MC simulations, we determined (see Ref.~\cite{CarpetIC}) the
following conditions for an event to be considered as candidate for a
$E_{\gamma}>1$~PeV photon: $N_{e} \ge N_{e}^{0}$, where $N_{e}^{0}$ is
chosen in such a way that this condition selects 95\% of
$E_{\gamma}>1$~PeV MC photons; $n_{\mu}/N_{e} \ge C$, where $C$ determines
the median of the distribution in $n_{\mu }/N_{e}$ among simulated photons
with $E_{\gamma}>1$~PeV. Figure~\ref{fig:cand-cuts}
presents these cuts on the $N_{e}-n_{\mu}$ plane with real events
superimposed; 523 events find themselves in the shaded area and are
considered as photon candidates. Though some or all of them might be
actually unusual muon-poor showers caused by primary hadrons (dedicated MC
simulations to determine the probability of such fluctuations are ongoing:
since we are discussing the very tail of the distribution, these
simulations require very large total number of generated events and a
careful account of the primary composition), their total number is
sufficiently low that any significant directional correlation with a
source would serve as an alert of their possible photon nature. Therefore,
we search for statistically significant excesses in the number of observed
photon candidates with arrival directions close to predefined sources. The
angle at which the correlation is searched for is determined as that
containing 90\% of reconstructed photons potentially coming from the
source. The observed number of events within this angle from the source is
then compared to that expected for isotropic distribution of arrival
directions (with the zenith-angle dependence of the efficiency taken into
account). The flux, or the upper limit on it, may then be determined with
the account of the direction-dependent exposure of the
experiment~\cite{CarpetIC}.
\section{Results}
\label{sec:results}
To avoid multiple tries and hardly controllable corrections to
significance, targets for the point-source search were determined before
looking at the data. They include: (i)~stacked arrival directions of
high-energy IceCube neutrino events; (ii)~individual arrival directions
from public IceCube alerts in the field of view of \textit{Carpet-2}, for
both directional and temporal correlations; (iii)~four predefined
astrophysical sources listed in Table~\ref{tab:limits}.
\begin{table}
\centering
\caption{Upper limits on the integral flux of gamma rays above 1~PeV from
four pre-defined sources. For Mrk~421, the best-fit flux is $7.6 \times
10^{-14}$~cm$^{-2}$\,s$^{-1}$.}
\label{tab:limits}
\begin{tabular}{cccc}
\hline
     & \multicolumn{2}{c}{photon candidates:} & Flux ($E_{\gamma}>$PeV),\\
Source& expected & observed & 95\% C.L.\ upper limit,\\
      &          &          & cm$^{-2}$\,s$^{-1}$\\
\hline
Crab &  0.4  & 0 & $2.1 \times 10^{-13}$ \\
Cyg~X-3 & 1.2  & 2 & $2.0 \times 10^{-13}$\\
Mrk~421 & 1.1 & 3& $2.7 \times 10^{-13}$\\
Mrk~501 & 1.1 & 1& $1.4 \times 10^{-13}$\\
\hline
\end{tabular}
\end{table}
The searches related to IceCube neutrinos are discussed in detail in
Ref.~\cite{CarpetIC}. We selected 34 published IceCube track-like events
with declinations in the field of view when  \textit{Carpet--2} was
taking data. Given individual geometry uncertainties of these events and
 the \textit{Carpet--2} angular resolution, 90\% of the signal is expected
 to be found within 3.0$^{\circ}$ from the directions. We found 10 photon
candidate events within 3.0$^{\circ}$ from these 34 IceCube arrival
directions, while 13.6 are expected from isotropy. This resulted in the
95\% CL upper limit on total steady flux of $E_{\gamma}>1$~PeV photons
from these 34 directions of $8.5\times 10^{-15}$cm$^{-2}$s$^{-1}$. In
addition, one event was in the  \textit{Carpet--2} field of view, a
$\approx 100$~TeV neutrino detected on December 10, 2016. Within a period
of 3 days centered on the neutrino arrival time, no candidate events
within 3.0$^{\circ}$ were observed while 0.02 are expected from isotropy;
the resulting 95\% CL upper limit on the fluence of the corresponding
flare in $E_{\gamma}>1$~PeV photons is $4.4\times 10^{-5}$~PeV/cm$^{2}$.

For astrophysical point sources, whose positions are known exactly, the
optimal search angle is just the 90\% CL angular resolution of the
experiment, 2.65$^{\circ}$. The results of the search are summarized in
Table~\ref{tab:limits}. A weak excess from the direction of Mrk~421 is
statistically consistent (2.4$\sigma$) with a fluctuation of the isotropic
background. We will monitor it in future data.
\section{Conclusions and outlook}
\label{sec:concl}
\begin{figure}
\centering
\sidecaption
\includegraphics[width=0.7\textwidth,clip]{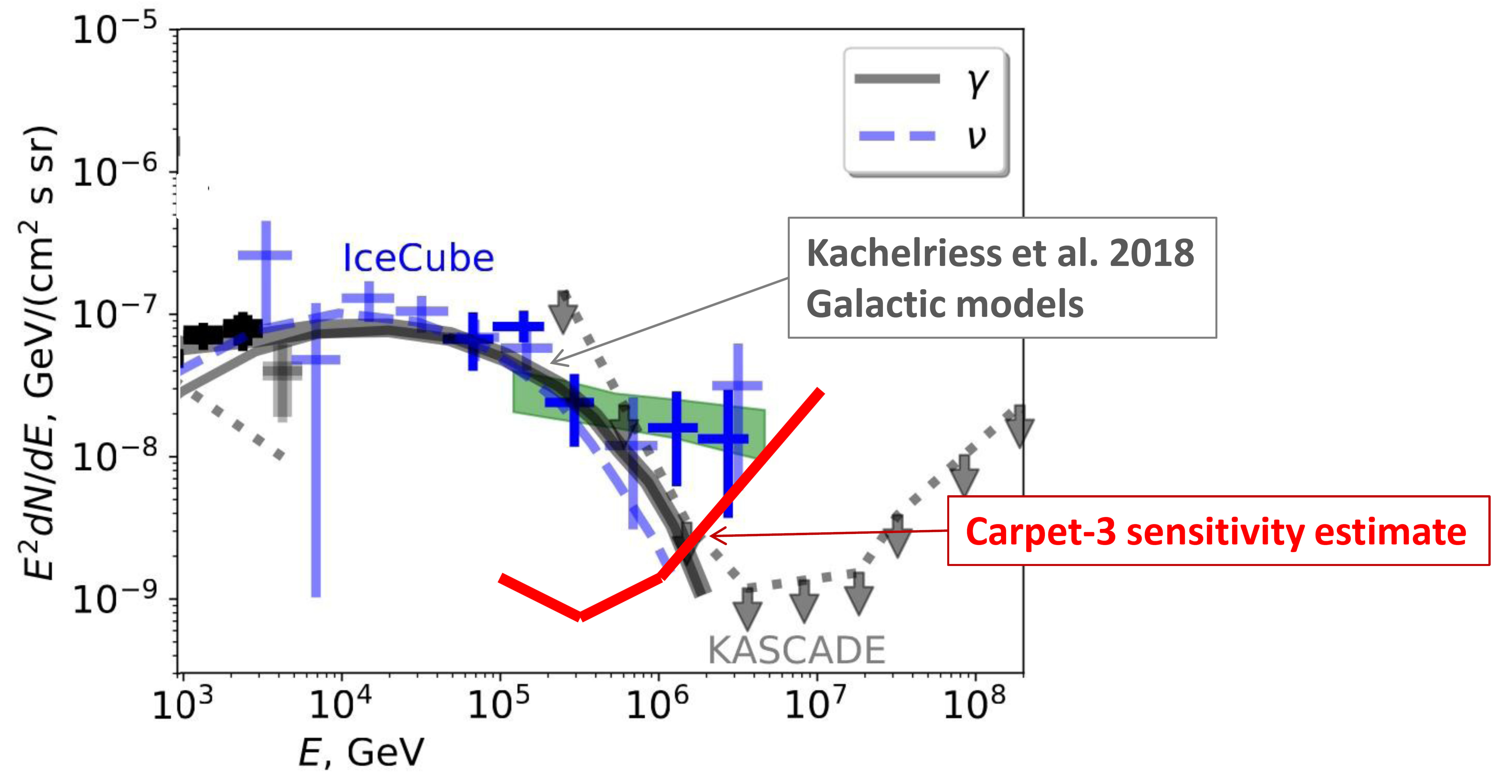}
\caption{%
Estimated Carpet--3 sensitivity (95\% CL, one year of live time) to the
diffuse gamma-ray flux (red line) superimposed on the plot with Galactic
model predictions from Ref.~\cite{Semikoz}.}
\label{fig:diffuse-plans}
\end{figure}

The results presented in this note are based on the data obtained with the
175~m$^{2}$ muon detector, shower cores in the central unit of 200~m$^{2}$
area and the trigger of $n_{\mu}>1$. In 2018, the installation started to
collect data with the new, ``photon-friendly'' trigger which accepts also
events with $n_{\mu}=0$ and 1. This would allow to lower the energy
threshold for photon searches. But a truly big step towards a better
gamma-ray sensitivity will be made in 2019, when it is planned that the
upgraded \textit{Carpet--3} installation will start taking data. Its
410~m$^{2}$ muon detector, already installed, will allow for a crucial
improvement in the gamma-hadron separation, while additional
surface-detector stations will increase the collecting area. Preliminary
MC estimates demonstrate that one year of \textit{Carpet--3} live data
taking would be sufficient to probe Galactic models of IceCube neutrinos
with the diffuse flux of $E_{\gamma}>100$~TeV photons, see
Fig.~\ref{fig:diffuse-plans}.
The corresponding improvement is expected for point sources as well.

\scriptsize
Experimental work of \textit{Carpet--2} is performed in the
laboratory of ``Unique Scientific Installation -- Baksan Underground
Scintillating Telescope''at the ``Collective Usage Center -- Baksan
Neutrino Observatory of INR RAS'' under support of the Program of
fundamental scientific research of the RAS Presidium ``Physics of
fundamental interactions and nuclear technologies''.
The work of a part of the group (DD, EG, MKh, NK, AUK, ANK, AL, OM, KP,
AY) on the upgrade of the installation was supported in part by the RFBR
grant 16-29-13049. The work of DD, ID, AUK, ANK, AL and VP was supported
in part by the RFBR grant 16-02-00687.
ST thanks Christian
Spiering for the invitation to the VLVNT-2018 workshop and Michael
Kachelrie\ss, Oleg Kalashev and Dmitri Semikoz for interesting
discussions. A part of computer calculations have been performed at the
cluster of INR Theoretical Physics Department.

\end{document}